\documentclass[12pt,a4paper,leqno]{amsart}
\usepackage{latexsym, amsfonts, amsthm, amsmath, amssymb, url, color, graphicx, epsfig, float, algorithm, algorithmic, program} 
\usepackage[mathscr]{euscript}
\newtheorem{teo}{Theorem}[section]

\usepackage{hyperref}

\usepackage[english]{babel}

\usepackage[mathscr]{euscript}

\makeatletter
\newif\ifmsbmloaded@

\newtheorem{defi}[teo]{Definition}

\input{scrload}

\title{Singularity of Sparse Circulant Matrices is NP-complete}
\author{Ilia Toli}
\address{\tt Northeastern University 360 Huntington Ave, Dept of Maths, \newline 540 Lake Hall, Boston, MA 02115-5000. ilia.toli@gmail.com.} 

\date{}
\begin{document}
\thispagestyle{empty}
\begin{abstract}
It is shown by Karp reduction that deciding the singularity of $(2^n - 1) \times (2^n - 1)$ sparse circulant matrices (SC problem) is NP-complete. We can write them only implicitly, by indicating values of the $2 + n(n + 1)/2$ eventually nonzero entries of the first row and can make all matrix operations with them. The positions are $0, 1, 2^{i} + 2^{j}$. The complexity parameter is $n$. Mulmuley's work on the rank of matrices \cite{Mulmuley87} makes SC stand alone in a list of 3,000 and growing NP-complete problems.
\end{abstract}

\keywords{NP-complete, circulant, matrix, singular, determinant, polynomial, equation} 

\maketitle

\section{Introduction}

It is proved that calculating the singularity of sparse circulant matrices (SC) is NP-complete. ''Sparse`` in our context means that only entries $0,1,2^i+2^j$ of the first row are eventually nonzero. The other rows are defined from the first one. In the next section we define 3 problems MQ, PQ, SC. MQ is a well established one. Our polynomial time reductions are $MQ \Rightarrow PQ \Rightarrow SC$. The reduction $MQ \Rightarrow PQ$ has already been established by Kipnis and Shamir (\cite{Kipnis:Shamir:99}, Section 3.) We give another proof in Appendix \ref{mqpq}. In the main body of this paper we deal with $PQ \Rightarrow SC$.

Most of the known 3000 and counting NP-complete problems are rewording of each other, iterative, or just variations of exhaustive search. Mulmuley \cite{Mulmuley87} has proven that calculating the rank of a matrix is none of the above, so SC holds a special place among all NP-complete problems. We can make nearly any (counterexamples exist) assumption on the shape of $n$ and the problem remains NP-complete. The commutative ring $(M_{k}, +, \cdot)$ of $k \times k$ matrices over a finite field is mathematically exceptionally rich. Matrices are very sparse, circulant, symmetric. We believe that SC has the potential to simplify the Complexity Zoo.

To the best of our knowledge, no previous work has been done in this direction. There exist very few NP-problems mentioning matrices, like MinRank, MaxRank, Sing \cite{Buss97thecomputational} but they are of a completely different nature. In MinRank e.g., we are given a matrix with part of the entries from a ring and part variables that assume values in that ring. The question is to find the values of the variables for which the matrix has the least rank. Here we are given all the entries of the matrix and ask whether the determinant is 0.

In the third section we prove that SC is NP-hard. In the fourth we provide SC with a certificate for {\tt yes} answers to any instance. This completes the NP-completeness proof. In fifth and sixth sections we give considerations on the present state of things and possible paths to follow. In appendix A we give an illustration of the algorithm of the fourth section. In appendix B we give another proof of $MQ \Rightarrow PQ$.

\section{Statement of the Problems}

Let's start by stating the two problems, one novel and one classic. We solve the classic NP-complete problem in polynomial time using as a subroutine the novel problem, and this proves the NP-hardness of the novel problem (Karp reduction.)

\subsection{MQ (Multivariate Quadratic)}

\begin{defi}
	A term is a monic monomial.
\end{defi}

\subsubsection{Instance} \cite{CGC-yesha} Polynomials $P_i[x_1, x_2, \dots x_n]$, $1 \leq i \leq m$, from the ring $\mathbb{F}_2[x_1,$ $x_2,$ $\dots,$ $x_n]$ of multivariate polynomials with coefficient in the field $\mathbb{Z}/2\mathbb{Z}$.

\subsubsection{Question} Do there exist $u_1, u_2, \dots, u_n \in \{0, 1 \}$ such that, for $1 \leq i \leq m$, $P_i[u_1, u_2, \dots u_n] = 0$?

\subsubsection{Comment} Remains NP-complete even if none of the polynomials has a term involving more than 2 variables, that is, all polynomials are quadratic. Easily solved in polynomial time if no term involves more than 1 variable, that is, all polynomials are linear, or if there is just one polynomial. Variant in which the $u_i$ are allowed to range over the algebraic closure of $\mathbb{F}_2$ is NP-hard, even if no term involves more than 2 variables. It is easy to prove that the problem remains NP-complete for all characteristics and if $m, n$ are polynomially related. By this case we can always assume that $m = n.$ If $m > n$, we can assume that $(m - n)$ variables are there and just don't appear. If $n > m$ we can count the last equation $(n - m)$ times.

\subsection{PQ (Pseudoquadratic)}

\begin{defi}
	A polynomial $P(x)$ as below with coefficients from some field $\mathbb{F}_q$ is called pseudoquadratic.
	\begin{equation} \label{ppp}
		P(x) = \sum_{i,j=0}^{n - 1} a_{ij}x^{2^i + 2^j} + \sum_{i=0}^{n - 1} b_{i}x^{2^i} + c
	\end{equation}
\end{defi}

In characteristic 2, $P(x)$ has $2 + n(n + 1)/2$ terms. It is rather sparse and writable in $O(n^2)$ space. The degree of $P(x)$ is at most, and generally equal to, $(2^{n - 1} + 2^{n - 2})$.

\subsubsection{Instance} (\cite{Kipnis:Shamir:99}, Section 3) A polynomial $P(x) \in \mathbb{F}_{2^n}[x]$ over the field $\mathbb{F}_{2^n}$.

\subsubsection{Question} Does there exist $u \in \mathbb{F}_{2^n}$ such that $P(u) = 0$?

\subsubsection{Comments} Transformation from MQ (\cite{Kipnis:Shamir:99}, Section 3). Remains NP-complete for a variety of generalizations that fall out of the scope of this paper. Not known to be NP-hard or NP-complete for $P(x)$ having just 3 terms. Trivially solved in polynomial time if it has 2 terms. It's trivial to bring MQ to PQ to MQ in polynomial time \cite{Kipnis:Shamir:99}.

\subsection{SC (Sparse Circulant)}

\begin{defi}
	A $n \times n$ circulant matrix is a matrix whose row number $k$ is obtained by left shifting the zeroth row $(a_{00}, a_{01}, a_{02}, \dots, a_{0,(n - 1)})$ by $k$ positions. That is, $a_{i, j} = a_{0, \{ (i + j) \mod n\} }$.
\end{defi}

The other equivalent definition is by shifting right. This one has the nice property that the matrix is symmetric. Two more equivalent definitions by columns.

\begin{defi}
	Here {\em a sparse circulant matrix} is a $(2^n - 1) \times (2^n - 1)$ circulant matrix with $2 + n(n + 1)/2$ nonzero entries in the first row, located in the positions 0, 1, $2^i + 2^j$.
\end{defi}

\subsubsection{Instance} A $(2^n - 1) \times (2^n - 1)$ sparse circulant matrix with entries from the finite field $\mathbb{F}_{2^n}$.

\subsubsection{Question} Is the matrix singular?

\subsubsection{Comment} Evidently, it takes exponential space in $n$ (the complexity parameter) to write the matrix explicitly. We assume writing it implicitly, by giving the nonzero entries of the first row. This is no hassle for performing the necessary matrix operations.

It is not known whether the problem remains NP-complete if the nonzero entries are just 3. Easily solved in polynomial time if the nonzero entries are 2.

\section{PQ implies SC, or SC is NP-hard}

We take it from Kipnis and Shamir (\cite{Kipnis:Shamir:99}, Section 3) that PQ is NP-complete. We bring PQ into SC in polynomial time by Karp reduction, that is, we'll solve PQ in polynomial time using SC as a subroutine.

We have
\begin{equation}
	\{ P(x) \mbox{ has roots in } \mathbb{F}_{2^n} \} \Leftrightarrow \{ \mbox{Res}(P(x), x^{2^n} - x) = 0 \}
\end{equation}
The Sylvester matrix associated with resultant Res$(P(x), x^{2^n} - x)$ is not circulant at all, we transform it into one with the same singularity.

\subsection{From Sylvester to circulant matrix.}

\begin{defi}
	Let ${\bf c} = (c_0, c_1, c_2, \dots c_{m-1})^t$ be a column vector and $f \neq 0$ a scalar. $Z_{f,m,n}(c) = (z_{ij})$ is called an $f$-circulant $m \times n$ matrix if $z_{ij} = c_{i-j \mod m}$ for $i \geq j$ and $z_{ij} = fc_{i-j \mod m}$ for $i < j$.

	For $f = 1$ it is circulant and for $f = -1$ anticirculant.
\end{defi}

For illustration of our steps, take $n = 3$, $\mathbb{F}_{2^n} = \mathbb{F}_{2^3}$ and a polynomial of degree $k < 2^n$ with coefficients in it, $p(x) = a_5x^5 + a_4x^4 + a_3x^3 + a^2x^2 + a_1x + a_0$, not necessarily sparse. Res$(x^{2^n} - x, p(x))$ is the determinant of the $(2^n + k) \times (2^n + k)$ Sylvester matrix in (\ref{sylvester}).

\begin{equation}  \label{sylvester}
 \left(
\begin{array}{ccccc|ccccccc|c}
1 & 0 & 0 & 0 & 0 & 0 & 0 & -1 & 0 & 0 & 0 & 0 & 0 \\
0 & 1 & 0 & 0 & 0 & 0 & 0 & 0 & -1 & 0 & 0 & 0 & 0 \\
0 & 0 & 1 & 0 & 0 & 0 & 0 & 0 & 0 & -1 & 0 & 0 & 0 \\
0 & 0 & 0 & 1 & 0 & 0 & 0 & 0 & 0 & 0 & -1 & 0 & 0 \\
0 & 0 & 0 & 0 & 1 & 0 & 0 & 0 & 0 & 0 & 0 & -1 & 0 \\
\hline
a_5 & a_4 & a_3 & a_2 & a_1 & a_0 & 0 & 0 & 0 & 0 & 0 & 0 & 0 \\
0 & a_5 & a_4 & a_3 & a_2 & a_1 & a_0 & 0 & 0 & 0 & 0 & 0 & 0 \\
0 & 0 & a_5 & a_4 & a_3 & a_2 & a_1 & a_0 & 0 & 0 & 0 & 0 & 0 \\
0 & 0 & 0 & a_5 & a_4 & a_3 & a_2 & a_1 & a_0 & 0 & 0 & 0 & 0 \\
0 & 0 & 0 & 0 & a_5 & a_4 & a_3 & a_2 & a_1 & a_0 & 0 & 0 & 0 \\
0 & 0 & 0 & 0 & 0 & a_5 & a_4 & a_3 & a_2 & a_1 & a_0 & 0 & 0 \\
0 & 0 & 0 & 0 & 0 & 0 & a_5 & a_4 & a_3 & a_2 & a_1 & a_0 & 0 \\
\hline
0 & 0 & 0 & 0 & 0 & 0 & 0 & a_5 & a_4 & a_3 & a_2 & a_1 & a_0 \\
\end{array}
\right)
\end{equation}

The case $a_0 = 0$ is trivial, assume $a_0 \neq 0$. The last column has only one nonzero entry. The singularity does not change after erasing the last column and row.

In (\ref{sylvester}) add column $i$ to column $(2^n + i)$ for $1 \leq i \leq k$, and get (\ref{add}).

\begin{equation} \label{add}
\left(
\begin{array}{ccccc|ccccccc}
1 & 0 & 0 & 0 & 0 & 0 & 0 & 0 & 0 & 0 & 0 & 0 \\
0 & 1 & 0 & 0 & 0 & 0 & 0 & 0 & 0 & 0 & 0 & 0 \\
0 & 0 & 1 & 0 & 0 & 0 & 0 & 0 & 0 & 0 & 0 & 0 \\
0 & 0 & 0 & 1 & 0 & 0 & 0 & 0 & 0 & 0 & 0 & 0 \\
0 & 0 & 0 & 0 & 1 & 0 & 0 & 0 & 0 & 0 & 0 & 0 \\
\hline
a_5 & a_4 & a_3 & a_2 & a_1 & a_0 & 0 & a_5 & a_4 & a_3 & a_2 & a_1 \\
0 & a_5 & a_4 & a_3 & a_2 & a_1 & a_0 & 0 & a_5 & a_4 & a_3 & a_2 \\
0 & 0 & a_5 & a_4 & a_3 & a_2 & a_1 & a_0 & 0 & a_5 & a_4 & a_3 \\
0 & 0 & 0 & a_5 & a_4 & a_3 & a_2 & a_1 & a_0 & 0 & a_5 & a_4 \\
0 & 0 & 0 & 0 & a_5 & a_4 & a_3 & a_2 & a_1 & a_0 & 0 & a_5 \\
0 & 0 & 0 & 0 & 0 & a_5 & a_4 & a_3 & a_2 & a_1 & a_0 & 0 \\
0 & 0 & 0 & 0 & 0 & 0 & a_5 & a_4 & a_3 & a_2 & a_1 & a_0
\end{array}
\right)
\end{equation}

In the first $k$ rows of (\ref{add}) there is only one nonzero entry per row. Singularity does not change after erasing the first $k$ rows and columns. We're left to find the singularity of the $(2^n - 1) \times (2^n - 1)$ circulant matrix in the right down square of (\ref{add}). This proves that SC is NP-hard. In order to finish the proof we must prove that given a {\tt yes} instance of SC (the determinant is 0) we can provide a proof, verifiable in polynomial time.

\section{The {\tt yes} Answer is Verifiable in Polynomial Time, or SC is NP-complete}

To a system $S$ of $n$ quadratic equations in $n$ variables we have associated the corresponding pseudoquadratic polynomial $P$, and to it the corresponding sparse circulant matrix $M$ (\ref{add}). We'll write $P_S(x), M_P$ and so on for the associated objects.

\begin{equation} \label{equivs}
(\mid \! M \! \mid \ = 0) \Leftrightarrow (P_M \mbox{ has solutions}) \Leftrightarrow (S_M \mbox{ has solutions}))
\end{equation}

If we have a subroutine that finds out that $\mid \! M \! \mid \ = 0$, we are able to recover a solution (all solutions if their number is polynomially bounded) of the associated polynomial $P_M(x)$ and of the associated system $S_M$. Each of these solutions works as a certificate: ''$\mid \! M \! \mid \ = 0$ because $P_M(x_0) = 0$.`` A solution to $P_M(x)$ can be verified in polynomial time. Same is true for a solution of $S_M$. We'll use $S_M$ in the generation of the certificate because it's technically easier.

It seems a little bit odd to work with $M$ and then for the proof of singularity regress to the problems we started from, but why not? It might be easier to guess the singularity of $M$ rather than guess whether $S_M$ has solutions, and then return to $S_M$ for the certificate. Next we give an algorithm for recovering all solutions of $S_M$ in polynomial time using as a subroutine the singularity of $M$ and ignoring the complexity of $M$.

\subsection{Producing a certificate}

All we're left to do is produce a solution to $S_M$ given $\mid \! M \! \mid \ = 0$. Recall from (\ref{equivs}) that $S_M$ has a solution iff $\mid \! M \! \mid \ = 0$.

By an argument of padding with zeros accordingly as in appendix \ref{mqpq} we can assume to have as many equations as variables in all systems we'll deal with, though the number of variables decreases by 1 per iteration and the number of equations typically remains constant.

We build a sparse binary tree whose nodes are systems derived from $S_M$. In this section drop the index $M$ for simplicity. The empty string is $\lambda$. The $k$-bit string $i = i_1 i_2 i_3 \dots i_k$ is the index of the system $S^k_i$ obtained from $S$ by substituting $x_{i_j} = i_j$. The children of $S_i^{k - 1}$ are $S_{i0}^k$ and $S_{i1}^k$. The depth of $S^k_i$ is $k$. The $ij$ means ''juxtapose strings $i$ and $j$.``

\begin{algorithm} \label{merlin}
\caption{Solving MQ using SC as a subroutine}
\begin{program}
k = 0 ; S^0_{\lambda} \ = S ; | check whether \ \ | \mid \! M_{S^0_{\lambda}} \! \mid | \ \ is \ 0, \ yes or no | ; \\ 
\WHILE k < n \DO
\FOR | every yes node \ \ | S_i^k  \ \ \DO
|write \ \ | S_{i0}^{k + 1}, \ \ \ S_{i1}^{k + 1}  ; \hspace{1.7cm}\COMMENT{construct the two children of $S_{i}^{k}$} \\
|check whether | \mid \! M_{S_{i0}^{k + 1}} \mid \! | and | \ \ \mid \! M_{S_{i1}^{k + 1}} \mid \!  \ \ | are \ 0, \ yes or no | ;
\OD ;
k++ ;
\OD ;
|return the indices of the yes nodes | ;
\end{program}
\end{algorithm}

Every {\tt yes} node in Algorithm (\ref{merlin}) will lead to at least one solution. It might bifurcate but cannot extinguish. If there are $s$ solutions, in any depth $k$ we can't have more than $s$ {\tt yes}-es. In the depth $(k + 1)$ we can't have more than $2s$ nodes altogether. This yields the non-sharp upper bound for the number of nodes, $2ns$.

Let's assume that writing down $S$ takes $O(n^{3})$ space. So, making one substitution is $O(n^{3})$. We need $2ns$ substitutions and the complexity becomes $O(sn^{4})$. At every substitution we have to check the singularity of the associated matrix. If $M$ is the complexity of checking once, \mbox{$O(sn^{4} + snM)$} is the overall complexity of the algorithm. We found all the solutions in polynomial time. This ends the proof of NP-completeness.

\section{Conclusions}

By the argument of padding with zeros as in appendix \ref{mqpq} we can make nearly every assumption about the shape of $n$ and the problem remains NP-complete. Experimentally our determinants do behave differently for different shapes of $n$ and this might be relevant. We've not yet been able to take advantage of it.

There are many ways to decide the singularity of sparse circulant matrices, the most promising of them all being Mulmulay's \cite{Mulmuley87}. Or we can check whether are there any zero eigenvalues. Calculating only one of them is rather inexpensive. There are $(2^n - 1)$ eigenvalues, checking them all is once exponential in $n$. We may check the number of linearly independent rows or columns. In general it comes down to Gaussian elimination, which here is exponential in $n$. We haven't yet taken any advantage of the special sparseness of our matrices and of other structure in them. Because $(2^n - 1)$ is odd, there always exist the $(2^n - 1)$th primitive roots of unity. Fast Fourier Transform methods apply. All these constructions work for all characteristics, with few adaptations.

It was proved that the problem of deciding the singularity of sparse circulant matrices is NP-complete. This is the first NP-complete problem of the kind. An even more interesting feature is their potential for further improvements of complexity.

None of algorithms mentioned here for calculating the singularity
\begin{itemize}
\item is novel or
\item takes advantage of
\begin{itemize}
\item the sparsity,
\item the being circulant and symmetric,
\item the fact that the ring of circulant matrices is commutative,
\item the shape of $n$,
\item deciding the singularity, as opposed to the determinant, rank or eigenvalues.
\end{itemize}
\end{itemize}
Further exploration of the topic is needed to clarify how much the complexity can be reduced with specially designed algorithms that take advantage from the very special structure, and whether that has any theoretical impact. 

\addcontentsline{toc}{section}{Bibliography}

\bibliographystyle{plain}


\appendix
\section{Example of Solving MQ in Polynomial Time Given SC}

For illustration of the Algorithm \ref{merlin} only, let's have the trivial system $S$ of 26 equations in 7 variables, which was generated in {\tt Singular} \cite{GPS}.

\begin{equation} \label{system}
\begin{array}{lllllll}
\begin{array}{l}
t_6t_7 \\
t_5t_7 + t_7 \\
t_4t_7 \\
t_3t_7 + t_7 \\
t_2t_7 + t_7 \\
t_1t_7 + t_7 \\
t_5t_6 + t_6 + t_5 + 1 \\
t_4t_6 + t_4 \\
t_3t_6 + t_3 + t_6 + 1
\end{array} & & &
\begin{array}{l}
t_2t_6 + t_6 + t_2 + 1 \\
t_1t_6 + t_6 + t_1 + 1 \\
t_5 + t_6 + 1 \\
t_4t_5 \\
t_3t_5 + t_5 \\
t_2t_5 + t_6 + 1 \\
t_1t_5 + t_6 + 1 \\
t_4 \\
t_3t_4 + t_4
\end{array} & & &
\begin{array}{l}
t_2t_4 \\
t_1t_4 \\
t_3 + 1 \\
t_2t_3 + t_2 \\
t_1t_3 + t_1 \\
t_2 + t_6 + 1 \\
t_1t_2 + t_6 + 1 \\
t_1 + t_6 + 1
\end{array}
\end{array}
\end{equation}

\

The execution of Algorithm \ref{merlin} gives the tree in Figure \ref{tree}. The {\tt yes} nodes are marked with * and are easy to guess from the graph anyway. We find the solutions $0010010$, $1110100$ and $1110101$ over $\mathbb{F}_2$. Some solutions are multiple but it doesn't matter in this argument. The complete tree has $2^7 = 128$ leaves, here we have only $14$.

\if false
\begin{figure} \label{tree}
{\tiny
\Tree [.{*S_{\lambda}} [ .{*S_0} [ .{*S_{00} } {S_{000}} [ .{*S_{001}} [ .{*S_{0010}} [ .{*S_{00100}} {S_{001000}} [ .{*S_{001001}} {*S_{0010010}} {S_{0010011}} ]] {S_{00101}}  ] {S_{0011}}  ]] {S_{01}}  ][ .{*S_1} {S_{10} } [ .{*S_{11}} {S_{110}} [ .{*S_{111}} [ .{*S_{1110}} {S_{11100}} [ .{*S_{11101}} [ .{*S_{111010}} {*S_{1110100}} {*S_{1110101}} ] {S_{111011}} ]] {S_{1111}} ]]]]
}
\caption{The tree for solving system S.}
\end{figure}
\fi

\section{The Transformation From MQ to PQ} \label{mqpq}

Let's have a system $S$ of $m$ equations in $n$ variables. By giving to all variables $x_1$, $x_2$, $x_3$, $\dots$ $x_n$ all possible values we define a function $(\mathbb{F}_{2})^n \rightarrow (\mathbb{F}_{2})^m$. After eventually padding in the end either one of the strings $n$, $m$ bit long, we can assume that $m = n$. The old elements of $(\mathbb{F}_{2})^m$ identify canonically with the ones that end with $(n - m)$ zeros. If $n > m$, we can keep the function as it is, sending nothing to the new elements. If $n < m$ we add $(m - n)$ variables and no equations for them. Canonically the new elements go to the $(\mathbb{F}_{2})^m$ element where goes the element to whom they're extensions.

So, the system $S$ defines a function $(\mathbb{F}_{2})^n \rightarrow (\mathbb{F}_{2})^n$ which induces a function $\mathbb{F}_{2^n} \rightarrow \mathbb{F}_{2^n}$. Every function in a finite field $\mathbb{F}_{2^n}$ is alternatively defined by a polynomial from $\mathbb{F}_{2^n}[x]$. If it has $t$ monomials to which we know the terms but not the coefficients and $t$ is small, we need only $t$ evaluations to get it by interpolation.

We have put in one-to-one correspondence each system of quadratic equations with a univariate polynomial with the property that every string that is a solution to the system is also a solution to the univariate polynomial, first considered as element of $(\mathbb{F}_{2})^{n}$ and then as element of $\mathbb{F}_{2^{n}}$.

In this shift from multivariate systems to univariate polynomials, to the quadratic systems of equations correspond pseudoquadratic polynomials (\cite{Kipnis:Shamir:99}, Section 3), or polynomials of the form (\ref{ppp}).

An alternative and constructive proof of this fact comes from the process of expanding a given univariate polynomial into a system of $n$ equations in $n$ variables. It's obvious that the correspondence between univariate polynomials $P(x)$ over $\mathbb{F}_{2^n}$ and systems $S_P$ of $n$ equations in $n$ variables over $\mathbb{F}_{2}$ is one-to-one. when expanding a term $x^d$ of $P(x)$ we get as many linear factors as the Hamming weight of $d$. The biggest Hamming weight will be the degree of the system. (This argument holds only for base field $\mathbb{F}_2$.) On pseudoquadratic polynomials it is by definition $2$.

\end{document}